\documentclass[11pt]{article}
\usepackage{authblk}
\usepackage{cite}
\usepackage{amsmath}
\usepackage{amssymb}
\usepackage{geometry}
\usepackage{caption}
\usepackage{color}
\usepackage{float}
\usepackage{graphicx}
\usepackage{hyperref}
\usepackage{epstopdf}

\usepackage{amssymb}
\usepackage{amsmath}
\usepackage{geometry}
\usepackage{units}
\usepackage{graphicx}
\usepackage{amsfonts}
\usepackage{float}
\usepackage{bm}
\usepackage{color}
\usepackage{epstopdf}
\usepackage{subcaption}
\usepackage{cite}
\title{A massless scalar particle coupled to the Wahlquist metric}

\author[1]{T. Birkandan \footnote{E-mail: birkandant@itu.edu.tr}}
\affil[1]{Istanbul Technical University, Department of Physics, Istanbul, Turkey.}

\author[2]{M. Horta\c{c}su \footnote{E-mail: hortacsu@itu.edu.tr}}
\affil[2]{Mimar Sinan Fine Arts University, Department of Physics, Istanbul, Turkey.}

\begin{document}
\maketitle

\begin{abstract}
We study the solutions of the wave equation where a
massless scalar field is coupled to the Wahlquist metric, a type-D
solution. We first take the full metric, and then write
simplifications of the metric by taking some of the constants in
the metric null. When we do not equate any of the arbitrary
constants in the metric to zero, we find the solution is given in
terms of the general Heun function, apart from some simple
functions multiplying this solution. This is also true, if we
equate one of the constants $Q_0$ or $a_1$ to zero. When both the
NUT related constant $a_1$ and $Q_0$ are zero, the singly
confluent Heun function is the solution. When we also equate the
constant $\nu_0$ to zero, we get the double confluent Heun-type
solution. In the latter two cases, we have an exponential and two
monomials raised to powers multiplying the Heun type function.
Thus, we generalize the Batic et al. result  for type-D metrics
for this metric and show that all variations of the Wahlquist
metric give Heun type solutions. 

\vspace{.5cm}

\end{abstract}

\section{Introduction}
\label{intro}

The Wahlquist metric, written as ``an exact interior solution for the finite rotating body of perfect fluid" was discovered in
1968 \cite {Wahlquist,Wahl1,Wahl2}. It is an axially symmetric, stationary, type-D solution of Einstein's field equation. Quoting
Wahlquist \cite {Wahlquist}, it can be  ``described as a superposition of a Kerr-NUT metric  \cite{Kerr,Newman} and a rigidly
rotation perfect fluid in the same space-time region."
The original metric written in \cite{Wahlquist}, was slightly modified by Senovilla \cite{Jose,Jose1}, and put to new form by Mars,
``to show that the Kerr-de Sitter and Kerr metrics are contained as subcases" \cite {Mars}.

Mars states in \cite {Mars} that Kramer \cite{Kramer1,Kramer2} showed the vanishing of the Simon tensor \cite{Simon} for this metric. Mars also states that the space-time admits a Killing tensor as shown in \cite {Papa}.
More recent work in this field exists, ``where the existence of a rank-2 generalized closed conformal Killing-Yano tensor with a
skew-symmetric torsion" \cite{Houri1}, and ``the separability of the Maxwell equation on the Wahlquist spacetime" are shown \cite{Houri2}.

Batic and Schmid showed, in their paper \cite{Batic}, that the solutions to the ``Teukolsky Master Equation" \cite{Teukolsky} could be
transformed in any physically relevant D type metric into one of the solutions of the Heun function, to its general or to one of its confluent
forms \cite {Ronveaux,Slavyanov,fiziev,Hortacsu,Tolga}. Here we check if this is true in the Wahlquist metric.

Another peculiarity of this metric is that, if we use Euclidean variables, the radial and angular equations look exactly like each other. We,
therefore, solve just the radial differential equation and get the solutions of the angular equation at the same time. This is a property shared by
the solutions in the background of the Kerr and Kerr-de Sitter  metrics  \cite{Kerr}, at least in a limit \cite{Gibbons}. The relation between the
 Wahlquist and Kerr solutions is generally known \cite{Wahlquist,Mars}.
It is interesting that an unrelated exact solution also share this property \cite{Yavuz,Yavuz1}.

Here we try to get the form of the exact solutions when a massless scalar field is coupled using this metric in full and also, in some cases when the
metric is reduced to a simpler form.
If all the constants, $Q_0$, $a_1$, $\beta$, $\nu_0$, $\mu_0$ appearing in the metric given in \cite{Mars} are not equal to zero, we get a general Heun solution
(HG) with four regular singularities multiplied by an exponential and  two polynomials. This is also true if we take just one constant in the
Wahlquist metric, $Q_{0} $, or $a_1$ null, and give arbitrary non zero values to the other constants. As explained below $lnx$ is approximated by
simple polynomials in this calculation.

If we set both $Q_0$ and the NUT parameter $a_1$ \cite{Newman1} equal to zero, we get the solution in terms of the (singly) confluent Heun (HC)
solution up to factors multiplying HC, as described above.
If we now equate still another constant, $\nu_{0}$, to zero, we get the double confluent Heun function (HD) up to multiplying exponent and mononomials
raised to powers. In the next section, we will describe our results. In an appendix, we give a calculation which is distantly related to this metric.
\section {Heun-type solutions}
Here we try to calculate the massless scalar $\psi$ where it obeys the equation
\begin{equation}
\frac{1}{\sqrt{g}} \partial_{\mu} g^{\mu \nu}\sqrt{g} \partial_{\nu} \psi = 0,
\end{equation}
using the metric given in \cite{Wahlquist}. Here $g$ is the determinant of the metric coefficents $ g_{\mu \nu}$.
We write the metric as is given in \cite {Houri1}, which is equivalent to the one given in \cite{Mars}. Here the comoving, pseudoconfocal,
spatial coordinates are used, which are closely related to the oblate-spheroidal coordinates in Euclidean geometry. We take $4 \pi G$ and $c$
equal to unity.
\begin{eqnarray}
ds^2= (v_{1} +v_{2}) \bigg(\frac{dz^2}{U}+ \frac{dw^2}{V} \bigg)
 +\frac{U}{v_{1} +v_{2}}(d\tau+v_{2}d\sigma)^2-\frac{V}{v_{1} +v_{2}}
(d\tau-v_{1}d\sigma)^2,
\end{eqnarray}
where
\begin{eqnarray}
U=Q_0+a_1{\frac{sinh(2\beta z)}{2\beta}}
-\frac{\nu_0}{\beta^2}{\frac{cosh(2\beta z)-1}{2\beta^2}}
-{\frac{\mu_0}{2\beta^2}}\bigg[{\frac{cosh(2\beta z)-1}{2\beta^2}} -z{\frac{sinh(2\beta z)}{2\beta}}\bigg],
\end{eqnarray}
\begin{eqnarray}
V=Q_0+a_2{\frac{sin(2\beta w)}{2\beta}}
+\frac{\nu_0}{\beta^2}{\frac{-cos(2\beta w)+1}{2\beta^2}}
-{\frac{\mu_0}{2\beta^2}}\bigg[{\frac{cos(2\beta w)-1}{2\beta^2}} +w{\frac{sin(2\beta w)}{2\beta}}\bigg],
\end{eqnarray}
and
\begin{equation}
v_1=\frac{cosh(2\beta z)-1}{2\beta^2}, \hspace{7pt} v_2=\frac{-cos(2\beta w)+1}{2\beta^2}.
\end{equation}
This metric has six real constants $Q_0,a_1,a_2,\nu_0,\mu_0,\beta$.
Here $a_1$ is related to the NUT \cite{Newman1,Halilsoy} parameter and $a_2$ is related to the mass parameter. One writes the other variables to
express the energy density, pressure and the fluid velocity of the perfect fluid. $ \beta$ is related to the scaling of $z$ and $w$, both space coordinates, in the linear transformation of these variables given in the original paper by Wahlquist \cite{Wahlquist}. $U$ and $V$ are related to $h_2, h_1$ in the original metric and to the invariant $\mu_0$, $\nu_0$. All these parameters are scaled so that they do not vanish in the $\beta$ going to zero limit.
The wave equation written in this metric separates easily. Our ansatz for the solution is
\begin{equation}
\psi = R(x) Y(w) T(\tau) S(\sigma).
\end{equation}
We have two Killing vectors since the metric does not depend on  $\tau$ and $\sigma$ explicitly, related to $t$ and $\theta$ in the original metric
 \cite{Wahlquist}. If we make a Wick rotation which changes $w$ to $y=iw$, and  $a_2$ to $-ia_2$ \cite  {Mars},  where $i$ is the square root of minus
 unity, the metric becomes symmetrical
\begin{eqnarray}
ds^2= (v_{1} +v_{2}) \bigg(\frac{dz^2}{U}- \frac{dy^2}{V} \bigg)
 +\frac{U}{v_{1} +v_{2}}(d\tau-v_{2}d\sigma)^2-\frac{V}{v_{1} +v_{2}}
(d\tau+v_{1}d\sigma)^2.
 \end{eqnarray}
Then, we have identical equations for $z$ and $y$. The equation for $z$ reads
\begin{eqnarray}
\partial_{z}(U \partial_{z}) R(z)T(\tau) S(\sigma) + \bigg(\frac{v_1^2}{U}\partial_{\tau}^{2}-2\frac{v_1 }{U}\partial_{\tau}\partial_
{\sigma}+\frac{1}{U} \partial_{\sigma}^{2}\bigg) R(z)T(\tau) S(\sigma)=0.
\end{eqnarray}
We get exactly the same equation for the new variable $y$, with appropriate changes like $U$ going to $V$  and $v_1$ going to $v_2$.
\begin{eqnarray}
\partial_{y}(V \partial_{y}) Y(y)T(\tau) S(\sigma) + \bigg(\frac{v_2^2}{V}\partial_{\tau}^{2}-2\frac{v_2 }{V}
\partial_{\tau}\partial_{\sigma}+\frac{1}{V}
\partial_{\sigma}^{2}\bigg) Y(y)T(\tau) S(\sigma)=0.
\end{eqnarray}
Since the functions $v_1, U$ (similarly $v_2, V$) do not depend on $\tau$ and $\sigma$, the solutions for $\tau$ and $\sigma$ are just exponential
functions, giving us constants upon differentiation.
Here we will try to solve the differential equation for a massless scalar field, minimally coupled in the background of this metric.

We find that this is not an easy task. The presence of hyperbolic sine and cosine functions in the wave equation prevents us from using standard
methods. We change our variables as $x=exp(2\beta z)$ which makes it possible to write the hyperbolic sine and cosine functions in terms of powers of
$x$. $sinh( {2\beta z}) = 1/2[x-1/x]$, $ cosh({2\beta  z})-1=1/2[x+1/x-2]$. Then, however, exists  the relation  $z=ln{x}/(2\beta)$. The codes we have to solve differential equations analytically, does not recognize $ln{x}$ for analyzing the singularities. Thus, we can not give an exact solution for all values of the independent variable $x$. We can get solutions only in different patches, by using  polynomial expressions approximating $lnx$ in this region to give us an idea
 what the solution may be.

We use different independent variables to check the validity of this approach. Luckily at different points that we expanded $lnx$, our solutions were
of the same type. We also plot  some of the solutions around these points. We use the package given in \cite{symode2} to analyze the singularity
structure of our equations.

Upon the variable change from $z$ to  $x$, we get the new equation
\begin{eqnarray}\label{fulleqn}
\frac {d^2 R(x)}{dx^2}
+\bigg(\frac{{\frac{d U}{d x }}}{U} +\frac{1}{x}\bigg) \frac{d }{d x } R(x)+ \frac{1}{x^2 U^2}\big(\omega^2v_1^2-2v_1\omega s +s^2\big)
 R(x)=0.
\end{eqnarray}
Here we used  the  solution to the $\tau$ equation in the form $exp(-\omega \tau)$ and the solution for the $\sigma$ equation in the form $exp(-\sigma s)$. Note that our metric had Killing vectors for both $\tau$ and $\sigma$.
Here
\begin{eqnarray}
U=Q_0 + a_1\frac{(x-1/x)}{4\beta}-\bigg[\frac{\mu_0}{4\beta^4}+\frac{\nu_0}{4\beta^3} \bigg]\frac{(x-1)^2}{x}+ \frac{\mu_0}{4\beta^4}ln x \frac{(x^2-1)}{x},
\end{eqnarray}
\begin{equation}
v_1= \frac{(x-1)^2}{2x\beta^2}.
\end{equation}
From here on, we will define $\mu_0'=\frac{\mu_0}{(4\beta^4)}$ and $\nu_0'= \frac{\nu_0}{(4\beta^3)}$, $a'_1= a_1/\beta$ and use these new constants
in our equations.

Before applying approximations, we can analyze our equation
(\ref{fulleqn}) numerically to see its behavior. Note that our
differential equation has four singular points, at zero, one and
at two other points. Here the point when $x=0$ is for the original
independent variable $z$ going to minus infinity. We, therefore,
start our graph at a point starting from $x>1$.

We take $Q_0=0.1, a_1=1, \beta=1.0,\mu_0'=1.0, \nu_0'=4.0,
\omega=1.0, s=10.0$ and use the $4th$ order Runge-Kutta method as
explained in \cite{cocuklu}.
\begin{figure}[!hbt]
	\centering
	\scalebox{0.6}{\includegraphics{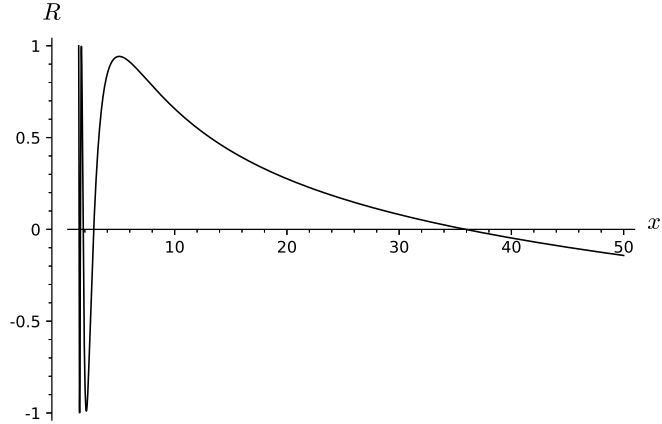}}
	\caption{The numerical behavior of Eq. (\ref{fulleqn}).}
	\label{fig:fig1}
\end{figure}

As we can see from the numerical solution given in Fig. (\ref{fig:fig1}) and from the function $U$ which governs the singularity behavior of the equation, the point $x=1$ is a singularity. Here, we start our numerical integrator very close to the singular point $x=1$ that yields an unstable behavior around this point. However, as we move towards $x>1$, we get a well-behaving curve.

To study the solution around the singularity $x=1$, we expand $lnx$ in the neighborhood of this point and use $x-1$ instead of $lnx$ in the wave equation we use in our further calculations.

We find that, if we keep all our constants non zero, we get the exact solution in terms of HG up to an exponential and terms (polynomials) multiplying
this function. The regular singularities are at $0$ and at three other finite points which are very lengthy expressions.

Then we go to the next possible choice. We try to find the solution when we keep all the terms in $U$ aside from $Q_0$, which we set equal to zero.
Then, our equation, again, has four regular singularities at $0,1$ and the points $-\frac{1}{2\mu'_0}(B \pm C^{1/2})$. Here
$B=(a'_1-\nu'_0-\mu'_0)$, $ C=B^2 - 4\mu'_0(a'_1+\nu'_0) $. It is known that this and the solution given in the paragraph above may be reduced to
 HG, the general Heun function \cite {Ronveaux,Slavyanov}.

We check this result by going to a different independent variable $u=\frac{(\mu_0'-\nu_0')x}{\nu_0' (1-x)}$, as suggested in the paper by
Suzuki et al. \cite {Suzuki}. For the general case, we
can identify the singular points. There are four regular singularities, at zero, infinity and two other finite points. This is consistent by the
result given above when the variable $x$ was used.

\begin{figure}[!hbt]
	\centering
\scalebox{0.6}{\includegraphics{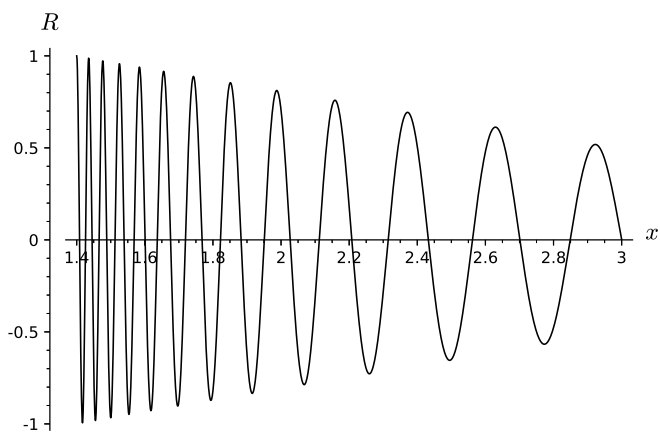}}
\caption{Numerical solution of Eq. (\ref{onuc}): $HC$ solution ($\beta=1.0,\mu_0'=1.0,\nu_0'=4.0,\omega=1.0,s=10.0$).}
  \label{fig:fig2}
\end{figure}

Then, we set $a'_1=0$, and try to see  what kind of  exact solution we obtain for this simplified case near the point $x=1$. We again approximate
$lnx$ by the expression $x-1$. Then, the equation reads

\begin{equation}{\label{onuc}}
\frac{d^{2} R(x)}{d x^{2}}+ \frac{2}{x } \frac{D}{ E}\frac{d R(x)}{d x}
+4\beta^2\frac{ F^2}{x^2 E^2 } R(x) =0,
\end{equation}
where
\begin{equation}
D=\mu'_0 x^2(x-1) -\frac{1}{2}(\mu'_0 + \nu'_0)(x-1)^2 - \frac{\nu'_0}{2}(x^2-1),
\end{equation}
\begin{eqnarray}
E=\mu'_0(x^2-1)(x-1)+(-\mu'_0-\nu'_0)(x-1)^2,
\end{eqnarray}
\begin{equation}
F= \omega x^2+ (4s-2\omega)x+\omega.
\end{equation}
Our solution  is a confluent Heun function, HC, which is multiplied by $exp(\frac{Asx}{x-1})$, and powers of $(x-\frac{\nu'_0}{\mu'_0})$ and $(x-1)$.
 Here the regular singular points are at $0, \frac{\nu'_0}{\mu'_0}$. $A$ is a constant. There is also an irregular singularity when $x$ equals unity.
 The numerical behavior of this equation is given in Fig. (\ref{fig:fig2}) with some numerical values of the physical parameters.

We again check this result by going to the independent variable $u=\frac{(\mu'_0-\nu'_0)x}{\nu'_0 (1-x)}$, to see if there is a change in the result.
We try to find the solution as $u$ goes to infinity. There we approximate
$lnx=ln \frac{u\nu'_0}{(u-1)\nu'_0+\mu'_0} $ by the expression  $\frac{(\nu'_0-\mu'_0)}{\nu'_0 u}$. The solution is in the same form. There are still
two regular singularities at $u= -\frac{\mu'_0}{\nu'_0}$ and
$u= -\frac{\mu'_0-\nu'_0}{\nu'_0}$, and  one irregular singularity at $u$ going to infinity. One of the regular singularities corresponds to $x$
equal to infinity and the irregular singularity is at $x$ equal to unity, where we had an irregular singularity when we used the variable $x$. The
existence of the confluent Heun function and the positions of these two singularities are consistent  with the singularities with the case when we
took $ln (x)$ as $x-1$.

\begin{figure}[!hbt]
	\centering
\scalebox{0.6}{ \includegraphics{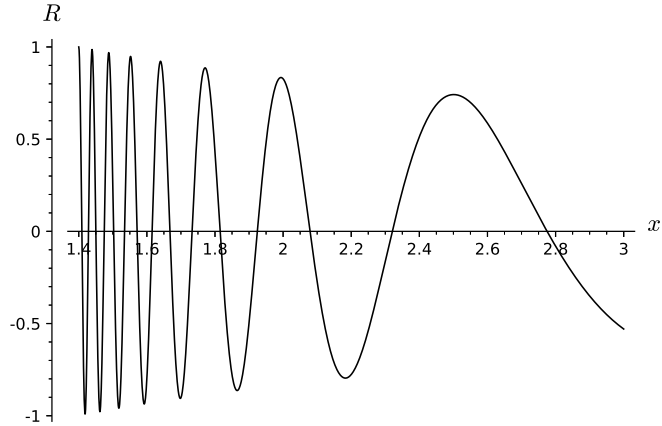}}
    \caption{The numerical behavior of Eq. (\ref{HDeqn}): $HD$ solution ($\beta=1.0,\mu_0'=1.0,\omega=1.0,s=10.0$).}
 \label{fig:fig3}
\end{figure}

As a final attempt, we try to find if we can get HD, double confluent Heun function \cite{Ronveaux}, from this solution. We
equate two regular singularities we found above. This may be obtained if we take the constant $\nu_0$ equal to zero. Since we divide by $\nu_0$ in some
of our latter expressions, we equate $\nu_0$ to zero in the original wave equation. This simplifies the wave equation. It reads
\begin{eqnarray}\label{HDeqn}
\frac{d^{2} R(x)}{d x^{2}} +2\bigg(\frac{1}{x}+\frac{1}{x-1} \bigg) \frac {d R(x)}{d x } + \bigg(\frac{\beta}{\mu'_0} \bigg)^2 \bigg( \frac{ \omega^2}{ x^4}+ \frac{8s \omega}{ x^3 (x-1)^2} + \frac{16s^2}{x^2 (x-1)^4}\bigg) R(x)=0.
\end{eqnarray}
In this expression, we approximated $lnx$ by $x-1$ to find the singularity near $x$ equal to unity. The irregular singularities $x=0$ and $x=1$ are
seen explicitly. There are no singularities at other points, including infinity. The solution will be HD function multiplied by exponentials. We need
them to get rid of the most violent singularities to get a solution as a power series expansion, as in HD. The numerical behavior of this equation is
given in Fig. (\ref{fig:fig3}).

\begin{figure}[!hbt]
	\centering
\scalebox{0.6}{ \includegraphics{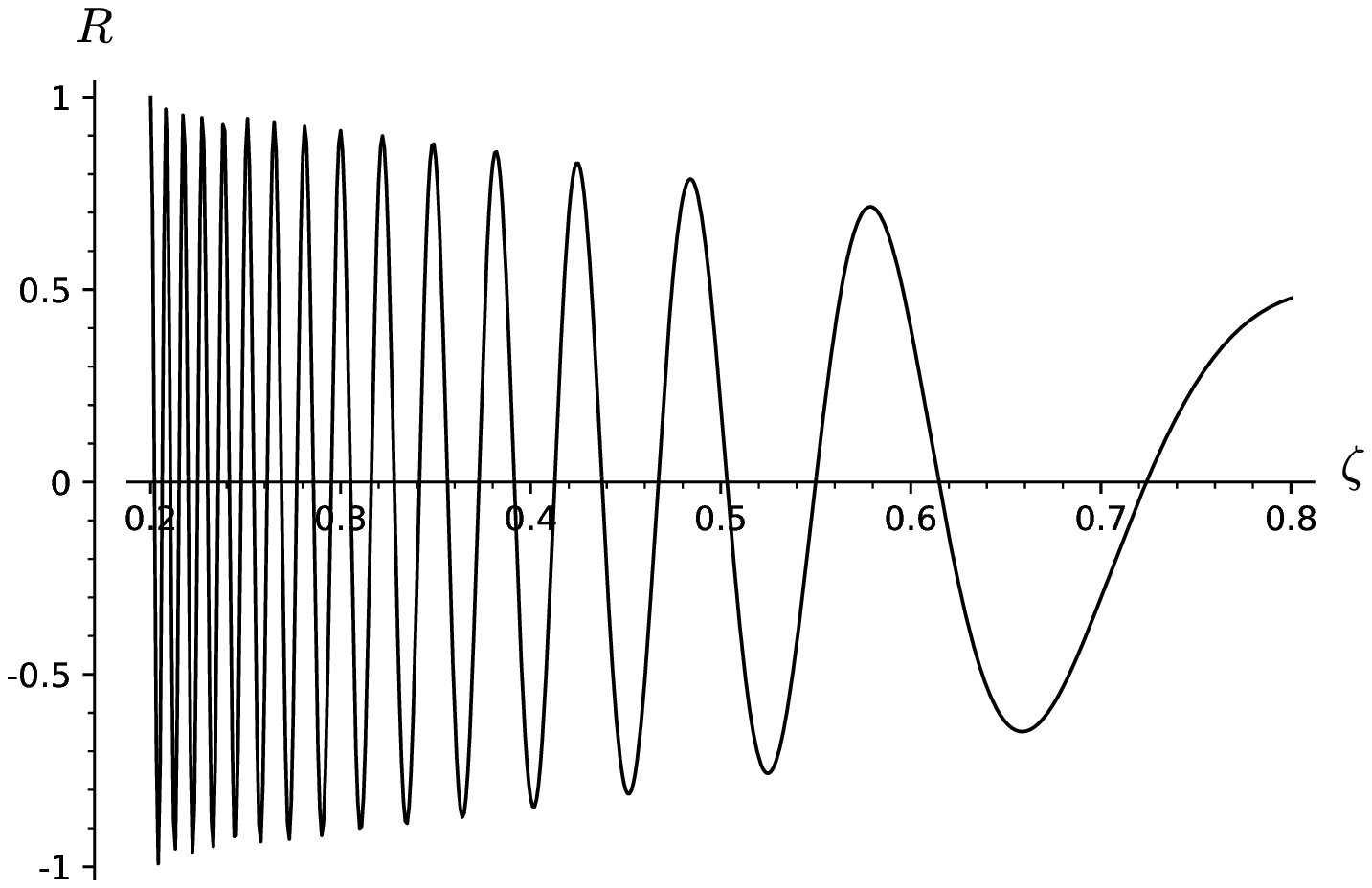}}
    \caption{The numerical behavior of the Eq. (\ref{zetaeqn}) ($\beta=1.0,\mu_0'=1.0,\omega=1.0,s=10.0$).}
    \label{fig:fig4}
\end{figure}

We again go to similar coordinates given in \cite {Houri1}, $\zeta=\frac{x-1}{x}$. Then the wave equation is written as
\begin{eqnarray}\label{zetaeqn}
\frac{d^{2} R(\zeta)}{d \zeta^{2}} +2\bigg(\frac{1}{1-\zeta}+\frac{1}{\zeta} \bigg) \frac{d R(\zeta)}{d \zeta}+
\bigg(\frac{\beta}{\mu'_0} \bigg)^2 \bigg(  \omega^2+ \frac{8s \omega (\zeta-1)}{\zeta^2 } + \frac{16s^2(\zeta-1)^2}{\zeta^4}\bigg)
R(\zeta)=0,
\end{eqnarray}
which explicitly shows the irregular singularity at $\zeta=0$ ($x=1$). The numerical behavior of this equation is given in Fig. (\ref{fig:fig4}).

To check the second irregular singularity, we take $\zeta=-1/\xi$. Then the wave equation reads
\begin{eqnarray}
\frac{d^{2} R(\xi)}{d \xi^{2}} -2\bigg(\frac{1}{\xi+1} \bigg) \frac {d R(\xi)}{d \xi} + \bigg(\frac{\beta}{\mu'_0} \bigg)^2
 \bigg( \frac{ \omega^2}{\xi^4}+ \frac{8s \omega (\xi+1)}{\xi^3 } + \frac{16s^2(\xi+1)^2}{\xi^2}\bigg) R(\xi)=0.
\end{eqnarray}
One should note that this equation is really valid for $\xi$ around $\xi=-1$. We anticipate that the general behaviour of the equation does not change
 close to $\xi$ around zero.

We can not check the behaviour of $ln x$ near $x=0$. This singularity comes as a regular singularity from our transformation from $z$ to $x$ in the
equation without approximations. It corresponds to the point as $z$ goes to minus infinity. Note that $lnx$ as $x$ goes to zero, does not create an additional singularity, since $lnx$ appears  as $x lnx$  in our expressions, and $x lnx $ is zero in this limit. For the case when only $Q_0,a_1$ are equal null, it appears as a regular singularity, giving us HC type solutions.

\section{Conclusion} Batic et al. \cite{Batic} showed that the solutions of the  equations are Heun-type if a type-D metric is used as a
background. Here we investigated if this is true for the Wahlquist metric and calculated  the solutions of the wave equation for a massless scalar
field, in the background of the Wahlquist metric. We studied first the full metric and then some reductions of it. We found that in all the four cases
studied, the solutions were Heun-type, namely HG, HC and HD for these different cases, thus, generalizing the Batic et al. result for the scalar field case. In an Appendix, we
showed that the same is true in a radically reduced form of this metric, which may no longer be classified as related to  the Wahlquist metric, since
the dust is not present.

\section{Acknowledgement}
M. H. is grateful to Prof. Reyhan Kaya, without whom this paper
would not be complete. He also thanks Taygun Bulmu\c{s} and Dr. Hasan Tuncay \"{O}z\c{c}elik for technical assistance in the early phases of this work.
He also acknowledges very fruitful conversations with Prof. Avedis S. Hacinliyan and Prof. M. Nazmi Postacıo\u{g}lu. The work of M. H. is morally
 supported by the Science Academy, Istanbul, an NGO.

\section{Appendix}
Here we will give an application of the Mathieu function, which is obtained in an extreme reduction of the metric given in our equation (1) where we
take both $U$ and $V$ equal to $Q_0$, taking all the other constants null. This metric will not be a solution of any special case of the Wahlquist metric, since in this metric $\mu_0$ must be different from zero.
\noindent
Mathieu function is a special case of HD \cite{Ronveaux1}. A similar calculation to the one given below was given in \cite{Yavuz1}.
\noindent
Here we will try to calculate  the Green function for a massless scalar field coupled to this special metric with both $U$ and $V$ are
constants, i.e. we try to calculate $G$ where it obeys the equation
\begin{equation}
\frac {1}{\sqrt{g}}\partial_{\mu} g^{\mu \nu}  \sqrt{g} \partial_{\nu} G (x,x') = - \delta^{4} (x,x'),
\end{equation}
where $g$ is the determinant of the metric coefficients $g_{\mu \nu}$.
Here $x,x'$ are generic independent variables, with no connection to $x$ used in the main text. This metric is not flat. The Ricci scalar is
given by
\begin{equation}
R = -16\beta^4 \frac { \cosh (2\beta z) \cos (2\beta w)-1}{(-\cosh (2\beta z)+\cos (2\beta w) )^3 }.
\end{equation}
First we write the equation for the wave equation when the right hand side is null. Our ansatz for the wave  solution $\phi$ is
\begin{equation}
\phi= R(z) P(w) T(\tau) S(\sigma).
\end{equation}
We will write the separated wave equation in two parts.
\noindent
\begin{eqnarray}
\frac{\partial_{z}^2 R(z)}{R(z)}+\bigg(\frac{v_1^2}{Q_0^2}
\frac{\partial_{\tau}^{2} T(\tau)}{T(\tau)} +2\frac{v_1}{Q_0^2}\frac{\partial_\tau T(\tau)}{T(\tau)} \frac{\partial_\sigma S(\sigma)}{S(\sigma)}+ \frac{1}{Q_0^2 } \frac{\partial_\sigma^{2} S(\sigma)}{S}\bigg)-\lambda =0,
\end{eqnarray}
\begin{eqnarray}
\frac{\partial_{w}^2 P(w)}{P(w)} + \bigg(\frac{v_2^2}{Q_0^2}
\frac{\partial_{\tau}^{2} T(\tau)}{T(\tau)}-
2\frac{v_1}{Q_0^2}\frac{\partial_\tau T(\tau)}{T(\tau)} \frac{\partial_\sigma S(\sigma)}{S(\sigma)}+\frac{1}{Q_0^2 }
\frac{\partial_\sigma^{2} S(\sigma)}{S}\bigg)= -\lambda.
\end{eqnarray}

Recall that
\begin{equation}
v_1=\frac{cosh(2\beta z)-1}{2\beta^2},
v_2=\frac{-cos(2\beta w)+1}{2\beta^2}.
\end{equation}
We take $ T= \exp(-i\tau k_{\tau})$, $ P(w)=\exp(-i\sigma k_{\sigma})$. We see that we can not simplify the problem in full generality. We
take $k_{\tau} =k \cos ( \phi), k_{\sigma} =k \sin ( \phi)$. To simplify, we have to fix $\tan { \phi} = \frac{1}{2 \beta^2}$, which means we are
 confined to a single line on the $\tau-\sigma$ plane. Of course we could our solution completely independent of either $\tau$ or $\sigma$ to give
 a similar result.
  If we continue with our first choice,  we get
\begin{eqnarray}
4 \beta^4 k^2(v_1^2+ 2v_1+ 1) = k^2 cosh^2 {2\beta z}=1/2 k^2(cosh {4\beta z} +1),
\end{eqnarray}
\begin{eqnarray}
4\beta^4 k^2(v_2^2 - 2v_2 +1 ) = k^2cos^2 {2\beta y} = 1/2 k^2 (cos {4\beta y}  +1),
\end{eqnarray}
\noindent
\begin{equation}
\frac{d^2P(w)}{dw^2} -( k^2 cos^2 (2\beta z) - \lambda) P(w) =0,
\end{equation}
and
\begin{equation}
\frac{d^2R(z)}{dz^2} -( k^2 cosh^2 (2\beta z)) R(z)=\lambda R(z).
\end{equation}
Solutions of both of these equations are expressed in terms of Mathieu functions. We see in Fig. 3 and 4, in a more general (HD) case, that these
are oscillation solutions. We first take the angular equation. ``We are interested only in the periodic solutions of this equation with period $2\pi$. These solutions exist only for discrete values of the separation constant $\lambda$ and they are given by even and odd periodic Mathieu functions
$Se_n( cos (2\beta w))$  and  $Se_n( cos (2\beta w))$ \cite{Yavuz1,Morse}. When the constants $k$ go to zero these solutions reduce to trigometric
functions and the seperation constants go to the square of an integer. The solutions of the $z$ may be expressed in terms of Bessel-Mathieu functions
$Je_n(2\beta\omega, cosh( 2\beta z))$, $Jo_n(2\beta\omega, cosh (2\beta z))$ and Hankel like Mathieu functions
$He_n(2\beta\omega, cosh( 2\beta z))$, $Ho_n(2\beta\omega,cosh( 2\beta z))$ respectively \cite{Morse}.

We sum over the discrete values from zero to infinity for the
Mathieu functions using formulae in \cite{Morse,Yavuz1}.
We have to evaluate
\begin{equation}
G(x,x')=\frac{1}{(2\pi)} {\int_{0}^{\infty}k \,dk
  e^{ikY}} g_h (z,z',w,W').
\end{equation}
\noindent
Here
\small
\begin{eqnarray}
\frac{g_h (z,z',w,w')}{ 4\pi} =
  \sum_{n=1}^{\infty} [ \frac {Se_{n} (h, \cos
   ( 2\beta w))Se_{n} (h, \cos ({2\beta w}'))  }{Me_{n}(h)}A 
+\frac{So_n(h, \cos (2\beta w')) So_{n} (h, \cos {(2\beta w))}}{Mo_{n}(h)}    B ],
\end{eqnarray}
\begin{eqnarray}
A= [\theta( z-z') Je_{n}(h, \cosh( \beta z')) H_{n}(h, \cosh (\beta z))]
+ [\theta( z'-z) Je_{n}(h, \cosh (\beta z))H_{n}(h, \cosh( \beta z'))],
\end{eqnarray}
\begin{eqnarray}
B=[\theta( z-z') Je_{n}(h, cosh( \beta z'))H_{n}(h, cosh( \beta z))]
+[\theta( z'-z) Je_{n}(h, cosh( \beta z))H_{n}(h, cosh (\beta z'))].
\end{eqnarray}
\normalsize
$\Theta ( z'-z)$ is the Heavyside unit step function and
\begin{equation}
Me_n=\int_{0}^{2\pi} |Se_n|^2 \,d\theta, Mo_n=\int_{0}^{2\pi} |So_n|^2 \,d\theta,
\end{equation}
where we equated the variable $2\beta w$  to $\theta$,
are used to normalize the above sum.
\noindent
One can show that
\begin{equation}
\frac{g_h (z,z',w,w')}{\pi}= H_{0} ( kZ),
\end{equation}
where $H_{0}$ is the Hankel function and
\begin{eqnarray}
&&4Z^2 = \cosh^2 (2\beta z')- \sin^2(2\beta w) + \cosh^2 (2\beta z))\nonumber \\  &&
- \sin^2(2\beta w')-2\cosh (2\beta z')\cosh (2\beta z) cos(2\beta w)cos(2\beta w')  \nonumber \\  &&
-2 \sinh(2\beta z)\sinh(2\beta z)\sin(2\beta w) \sin(2\beta w).
\end{eqnarray}
\begin{equation}
Y^2= (\tau-\tau')^2+(\sigma-\sigma')^2,
\alpha=Y .
\end{equation}
We first perform the $k$ integration using the standard integrals \cite{Ryznik1}
\begin{equation}
\int_{0}^{\infty}k \exp(ik Y) J_0(kZ)\,dk = ( Z^2-Y^2)^{-1}
P_{1}(\frac{iY}{\sqrt{Z^2-Y^2}}),
\end{equation}
\begin{equation}
\int_{0}^{\infty}k \exp(ik Y) N_0(kZ)\,dk = ( Z^2-Y^2)^{-1}
Q_{1}(\frac{i}{ \sqrt{Z^2-Y^2}}).
\end{equation}
where $P_1$  and $Q_1$ are Legendre functions of first and second kind.
We can also use the formula \cite{Ryznik2}
\begin{eqnarray}
\int_{0}^{\infty}k \exp(ik Y) K_0(kZ)dk = ( -Z^2-Y^2)^{-1}
\Big(\frac{Y}{\sqrt{-Z^2 -Y^2}}\ln {\big[ Y/Z + \sqrt{\frac{Y^2}{Z^2}-1} \big] } -1 \Big),
\end{eqnarray}
which gives
\begin{eqnarray}
\int_{0}^{\infty}k \exp(ik Y) K_0(-ikZ)\,dk = ( Z^2-Y^2)^{-1}
\Big(\frac{iY}{\sqrt{Z^2 -Y^2}}\ln {\big[ Y/Z + \sqrt{
\frac{Y^2}{Z^2} -1 }\big]} -1 \Big).
\end{eqnarray}
This is the  result we obtain  for the Green's function in three-dimensions for this metric.

\end{document}